\documentclass[twocolumn,prl,aps,showpacs]{revtex4}
\usepackage{graphicx}

\begin{document}

\title{\bf Low-energy Effective Theory for Spin Dynamics of Fluctuating Stripes}

\author{Chi-Ho Cheng}
\email{phcch@phys.sinica.edu.tw} \affiliation{Institute of
Physics, Academia Sinica, Taipei, Taiwan} \affiliation{Institute
of Theoretical Physics, Chinese Academy of Sciences, Beijing}

\date{\today}

\begin{abstract}
We derive an effective Hamiltonian for spin dynamics of
fluctuating smectic stripes from the {\it t-J} model in the weak
coupling limit $t\gg J$. Besides the modulation of spin magnitude,
the high energy hopping term would induce a low-energy
anti-ferromagnetic interaction between two neighboring ``blocks of
spins". Based on the effective Hamiltonian, we applied the linear
spin-wave theory and found that the spin-wave velocity is almost
isotropic for $\rm La_{2-x}Sr_x CuO_4$ unless the structural
effect is considered. The intensity of the second harmonic mode is
found to be about 10\% to that of the fundamental mode.
\end{abstract}


\maketitle

\vspace{-0.10cm}

There's still a lot of interest on stripe physics
\cite{review,review2}. The stripe modulation was found in cuprate
superconductor \cite{tranquada} of the sample $\rm La_{2-x}Sr_x
CuO_4$ (LSCO) in the low-temperature orthorhombic (LTO) structure.
It is also consistent with the incommensurate magnetic peaks
observed in the inelastic neutron scattering experiments
\cite{incomm-expt}, in which the peak shift from $(\pi,\pi)$
towards $(\pi,0)$ with the derivation very near to $2\pi\delta$ at
doping concentration $\delta$. By partial substitution of Nd for
La, the LTO lattice structure is distorted to the low-temperature
tetragonal (LTT) structure \cite{crawford}, in which the
horizontal stripe is enhanced and the fluctuating (dynamic) stripe
becomes more ordered (static).

Theoretically, by the mean-field analysis of the single-band
Hubbard model \cite{hubbard}, there is a possibility that the
stripe phase is formed. On the other hand, by employing the
Schwinger-boson mean-field theory \cite{auerbach} to the {\it t-J}
model, it is found that the spiral spin state \cite{incomm-th} can
also give the deviation from $(\pi,\pi)$ upon doping. However, it
was studied by many approaches that the uniform phase is unstable
towards phase separation in the range of interest ratio {\it t/J}
\cite{phase-separation}. One of the consequence of the phase
separation is to form stripes, in order to be consistent with the
neutron scattering experiments. Because of the stripe fluctuation
around the hole domain, the spins across the hole domain should be
anti-parallel. By considering just a small transverse fluctuation
of the stripe, it was found that two neighboring spins across the
hole domain feels an anti-ferromagnetic interaction of coupling
$J$ \cite{zachar02}.


Although there is still controversy about the stability of stripes
\cite{stability}, in this paper, the fact that the fluctuating
smectic stripe phase is stable is our assumption.
By including the Coulomb repulsion (which is neglected in the {\it
t-J} model), the holes can only phase separate at microscopic
scale instead of full phase separation. In order to balance the
hole kinetic energy and the Coulomb repulsion, stripes is the
simplest solution among those inhomogeneous states at microscopic
scale.

In the following, we are going to derive the low-energy spin
dynamics from the 2D {\it t-J} model of the site-centered stripe
along $y$-direction. The period of hole domain is $2R$ where $R\gg
1$ in our model.

The hopping term $H_t$ in the {\it t-J} model is written as
\begin{eqnarray}
H_t &=& H_t^{\perp} + H_t^{\parallel} \nonumber \\
 &=& \sum_\alpha  \left( H_t^{\perp,\alpha} +
 H_t^{\parallel,\alpha} \right)
\end{eqnarray}
where $H_t^{\perp}$ is the transverse fluctuation of vertical
stripes, and $H_t^{\parallel}$ is the 1D kinetic motion along the
vertical stripes. Now
\begin{eqnarray} \label{htalpha}
 H_t^{\perp,\alpha} &=& -t \sum_{\sigma} {\sum_{i}}'
 \left(
 c_{i+\hat x,\alpha,\sigma}^\dagger c_i + c_{i+2\hat x,\alpha,\sigma}^\dagger c_{i+\hat x,\alpha,\sigma} +
 \ldots  \right. \nonumber \\
 && \left. + c_{i+R\hat x,\alpha,\sigma}^\dagger c_{i+(R-1)\hat x,\alpha,\sigma}
  \right) + {\rm h.c.}
\end{eqnarray}
where $\alpha$ label the smectic stripes and $i_x$ is summed over
the super-cell of period $2R$. And also
\begin{eqnarray}
H_t^{\parallel,\alpha} = -t \sum_{\sigma} {\sum_{i}}'
 c_{i+\hat y,\alpha,\sigma}^\dagger c_{i,\alpha,\sigma}
 + {\rm h.c.}
\end{eqnarray}

In the weak coupling limit, $t \gg J$, the fluctuating stripe
induces a charge-density wave along $x$-direction. There's two
almost degenerate ground states for $H_t^{\perp,\alpha}$,
$|\psi_{\rm S}\rangle$ and $|\psi_{\rm T}\rangle$, which are
expressed as
\begin{eqnarray} \label{psi_st}
|\psi_{\rm S}\rangle &=& \frac{1}{\sqrt{2}} \left(
 |\psi_1\rangle - |\psi_2\rangle \right) \\
|\psi_{\rm T}\rangle &=& \frac{1}{\sqrt{2}} \left(
 |\psi_1\rangle + |\psi_2\rangle \right)
\end{eqnarray}
where
\begin{eqnarray}
|\psi_1\rangle &=&
 a_0 |\cdot\cdot \downarrow \uparrow \circ \downarrow \uparrow \cdot\cdot\rangle
  + a_1 |\cdot\cdot \downarrow \uparrow \downarrow \circ  \uparrow
  \cdot\cdot \rangle
   \nonumber \\
 && + a_{-1} |\cdot\cdot \downarrow  \circ \uparrow \downarrow \uparrow \cdot\cdot \rangle + \ldots \\
|\psi_2\rangle &=&
 a_0 |\cdot\cdot \uparrow \downarrow \circ \uparrow \downarrow
 \cdot\cdot\rangle
 + a_1 |\cdot\cdot \uparrow \downarrow \uparrow \circ  \downarrow
  \cdot\cdot \rangle
 \nonumber \\
 && + a_{-1} |\cdot\cdot \uparrow  \circ \downarrow \uparrow \downarrow \cdot\cdot \rangle
  + \ldots
\end{eqnarray}
The basis of $|\psi_1\rangle$ and $|\psi_2\rangle$ are different
by the up and down spins. The corresponding coefficients are
equal. Two subspaces $\{|\psi_1\rangle\}$ and $\{|\psi_2\rangle\}$
are orthogonal to each other. By Eq.(\ref{psi_st}), which is an
orthogonal transformation from $|\psi_1\rangle$ and
$|\psi_2\rangle$ to $|\psi_{\rm S}\rangle$ and $|\psi_{\rm
T}\rangle$, two subspaces ${\cal H}_S$ and ${\cal H}_T$ spanned by
$\{|\psi_{\rm S}\rangle\}$ and $\{|\psi_{\rm T}\rangle\}$
respectively are also orthogonal to each other. Eq.(\ref{htalpha})
can be written in the following matrix form under the above basis,
{\it i.e.},
\begin{eqnarray}
H_t^{\perp,\alpha} = (-t) \left(\begin{array}{ccccc}
 0 & 1 & 1 & 0 &  \ldots \\
 1 & 0 & 0 & 1 &  \ldots \\
 1 & 0 & 0 & 0 &  \ldots \\
 0 & 1 & 0 & 0 &  \ldots \\
 \vdots & \vdots & \vdots & \vdots & \ddots
 \end{array}\right) \bigotimes (-t)
 \left(\begin{array}{ccccc}
 0 & 1 & 1 & 0 & \ldots \\
 1 & 0 & 0 & 1 & \ldots \\
 1 & 0 & 0 & 0 & \ldots \\
 0 & 1 & 0 & 0 & \ldots \\
 \vdots & \vdots & \vdots & \vdots & \ddots
 \end{array}\right) \nonumber \\
\end{eqnarray}
where the two identical $(2R+1)$-dimensional matrices correspond
to two orthogonal subspaces spanned by $\{|\psi_{\rm S}\rangle\}$
and $\{|\psi_{\rm T}\rangle\}$. Simultaneously, the Heisenberg
term in the {\it t-J} model is written in the matrix form
\begin{eqnarray}
 H_J^{\perp,\alpha} = (-\frac{3J}{4}) \left(\begin{array}{ccccc}
 0 & 0 & 0 & 0 &  \ldots \\
 0 & 1 & 0 & 0 &  \ldots \\
 0 & 0 & 1 & 0 &  \ldots \\
 0 & 0 & 0 & 1 &  \ldots \\
 \vdots & \vdots & \vdots & \vdots & \ddots
 \end{array}\right) \bigotimes (\frac{J}{4}) \left(\begin{array}{ccccc}
 0 & 0 & 0 & 0 &  \ldots \\
 0 & 1 & 0 & 0 &  \ldots \\
 0 & 0 & 1 & 0 &  \ldots \\
 0 & 0 & 0 & 1 &  \ldots \\
 \vdots & \vdots & \vdots & \vdots & \ddots
 \end{array}\right) \nonumber \\
\end{eqnarray}

Now because two subspaces ${\cal H}_{\rm S}$ and ${\cal H}_{\rm
T}$ are orthogonal to each other, the first-order correction to
the ground state energy due to perturbed term $H_J^{\perp,\alpha}$
is
\begin{eqnarray} \label{ergs}
E_{\rm S}^{(1)} &=& \langle \psi_{\rm
S}|H_J^{\perp,\alpha}|\psi_{\rm S}\rangle + O(J^2/t)  \\
\label{ergt} E_{\rm T}^{(1)} &=& \langle \psi_{\rm
T}|H_J^{\perp,\alpha}|\psi_{\rm T}\rangle + O(J^2/t)
\end{eqnarray}
and no correction to the wavefunction since two subspaces are
orthogonal. Solving the eigenproblem for $H_t^{\perp,\alpha}$
gives the transverse density $\rho(x)$ in $(-R,R)$ to the leading
order can be obtained from the
\begin{eqnarray} \label{rho}
\rho(x) &=& |\langle x|\psi_{\rm S}\rangle|^2 = |\langle
x|\psi_{\rm T}\rangle|^2 \nonumber \\ &=& \frac{\delta}{1+2/\pi}
\left( 1+ \cos(\frac{\pi x}{2R}) \right)
\end{eqnarray}
and repeat for a period $2R$. It would be convenient to define
$\vec q_0 =(q_0,0)$, where $q_0 = \pi/(2R)$. The charge-density
wave induced by stripe fluctuation is
\begin{eqnarray}
\rho(\vec r) = \delta + \rho_1 \cos(2 \vec q_0 \cdot \vec r)
 + \rho_2 \cos(4 \vec q_0 \cdot \vec r) +
\ldots
\end{eqnarray}
where $\rho_1=\frac{4\delta}{3(2+\pi)}$ and
$\rho_2=-\frac{4\delta}{15(2+\pi)}$.

Eqs.(\ref{ergs})-(\ref{ergt}) tells that two almost degenerate
states are split by an energy difference $E_{\rm T}^{(1)}-E_{\rm
S}^{(1)}$, and hence the energy difference between two ``block of
spins" is
\begin{eqnarray}
J' = 2R \delta(E_{\rm T}^{(1)}-E_{\rm S}^{(1)})
 = \frac{2J \delta R^2 }{R+1}
\end{eqnarray}
Substitution of $q_0=2\pi\delta$ gives $J'=J/(2(1+4\delta))$.
Notice also that the energy difference per site between anti-phase
and in-phase becomes $J'/(2R)=J \delta /(1+4\delta)$, which is
almost linear in $\delta$ for $\delta\ll 1$. For $\delta = 1/8$,
energy difference is $J/12 \simeq 130K$ (take $J=135meV$), which
is consistent with the observation of the incommensurate peak up
to around $100K$ \cite{incomm-expt}.



\begin{figure}[tbh]
\begin{center}
\includegraphics[width=3in]{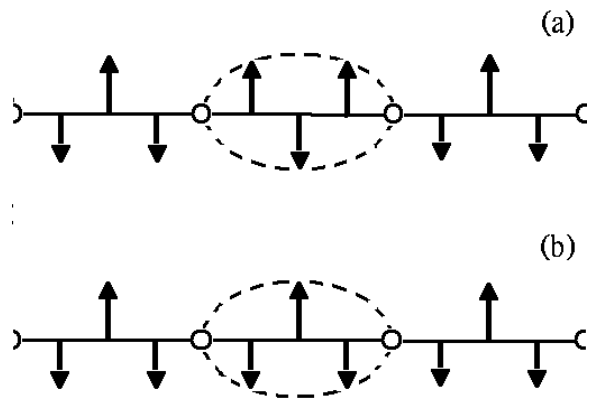}
\end{center}
\vspace{-5pt}
 \caption{Illustration of the flip of a ``block of spins" from (a) to (b).}
 \label{spin4.eps}
 \vspace{-5pt}
\end{figure}

There is then an effective anti-ferromagnetic coupling term
between two neighboring blocks of spins. The anti-phase described
by $|\psi_{\rm S}\rangle$ is of lower energy $J'$ than the
in-phase by $|\psi_{\rm T}\rangle$. Notice that it is different
from the case that two neighboring spins across the hole domain
interacts with an anti-ferromagnetic coupling
\cite{zachar02,otherHam}. The illustration of how to flip a block
of spins is shown in Fig.\ref{spin4.eps}. Several mean-field type
studies neglecting this low-energy interaction \cite{ogata} cannot
distinguish anti-phase and in-phase.

Because of no-double occupancy at every site, the spin magnitude
also form a period of $2R$, in which
\begin{eqnarray}
S(\vec r) &=& \frac{1}{2}(1-\rho(\vec r)) \nonumber \\ &=& S_0 +
S_1 \cos(2 \vec q_0 \cdot \vec r) + S_2 \cos(4 \vec q_0 \cdot \vec
r ) + \ldots
\end{eqnarray}
where $S_0 = \frac{1}{2}(1-\delta)$, $S_1 =
-\frac{2\delta}{3(2+\pi)}$, and $S_2=\frac{2\delta}{15(2+\pi)}$.

\begin{figure}[tbh]
\begin{center}
\includegraphics[width=3in]{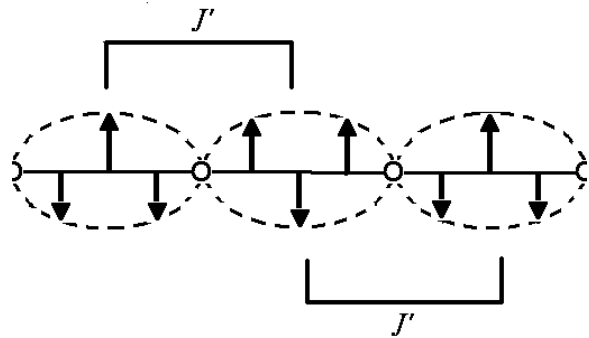}
\end{center}
\vspace{-5pt}
 \caption{Schematic diagram of spin state at doping concentration $\delta=1/8$
 of LSCO ($q_0 = 2\pi\delta = \pi/4$).
 The circles represent the zero average of spin direction. The dashed line shows the
 envelope of the spin component in $z$-direction. Any two neighboring ``blocks of spins" feel
 an anti-ferromagnetic interaction of coupling $J'$.}
 \label{spin2.eps}
\end{figure}

Because of the anti-ferromagnetic interaction between two blocks
of spins across the hole stripes, we can determine {\it
classically} the $z$-component of the spin $S^z(\vec r)$ which
gives
\begin{eqnarray} \label{sz}
S^z(\vec r) {\rm e}^{i\vec Q \cdot \vec r} = S^z_1 \sin(\vec q_0
\cdot \vec r ) + S^z_3 \sin(3 \vec q_0 \cdot \vec r) + \ldots
\end{eqnarray}
where $\vec Q = (\pi,\pi)$,
$S^z_1=\frac{2}{\pi}-\frac{3\delta}{2+\pi}$, and
$S^z_3=\frac{2}{3\pi}-\frac{5\delta}{3(2+\pi)}$. Notice that the
components $S^z_2$, $S^z_4$, and etc vanish because of the
anti-phase symmetry. Including the effect from lattice distortion
gives the same result. The non-vanishing higher harmonic should
have wavevector $3q_0$, $5q_0$, and etc. The schematic diagram of
the classical spin state is shown in Fig.\ref{spin2.eps}.


Estimated up to the order of magnitude, the neutron scattering
intensity is more or less proportional to
\begin{eqnarray}
&& \int d{\vec r} {\rm e}^{-\vec k \cdot \vec r}
 \langle S^z(\vec r) S^z(0) \rangle \nonumber \\
 &\propto& (S^z_1)^2 \left[\delta(\vec k-\vec Q+\vec q_0)+\delta(\vec k-\vec Q-\vec
 q_0)\right] \nonumber \\
 && + (S^z_3)^2 \left[\delta(\vec k-\vec Q+ 3 \vec q_0)+\delta(\vec k-\vec Q- 3\vec
 q_0)\right] \nonumber \\ && + \ldots
\end{eqnarray}
For $\delta\ll 1$, $(S^z_3/S^z_1)^2 \simeq 1/9$. Substitute
$\delta = 1/8$, the ratio is 0.093. In general, the ratio of the
intensities of the second harmonic ($\vec Q \pm 3\vec q_0$) to
that of  the fundamental mode ($\vec Q \pm \vec q_0$) is about
10\%. However, the signal-to-noise ratio in current neutron
scattering experiments \cite{tranquada} is not high enough to
observe the second harmonic peak. We expect further experiments on
a larger pure single crystal measurement can verify our
prediction. The first harmonic occuring at $\vec Q\pm 2\vec q_0$
vanishes. For LSCO, structural effect at doping concentration
$\delta=1/8$ enhances the fundamental mode so that the second
harmonic is even harder to be observed.

In order to obtain the low-energy properties, we apply the linear
spin-wave expansion \cite{hp}, {\it i.e.},
\begin{eqnarray}
\hat{S}^z(\vec r_i) &=& \pm (S(\vec r_i)-b_i^\dagger b_i) \\
\hat{S}^\pm(\vec r_i) &=& \sqrt{S(\vec r_i)} b_i \\
\hat{S}^\mp(\vec r_i) &=& \sqrt{S(\vec r_i)} b_i^\dagger
\end{eqnarray}
in which $b_i^\dagger$'s and $b_i$'s describe the quantum
fluctuation from the classical ground state. The effective
Hamiltonian for spin dynamics is
\begin{eqnarray}
H_J &=& E_0 + 4J\sum_i S(\vec r_i) b_i^\dagger b_i \nonumber \\ &&
+ 2J\sum_i \left( \sqrt{S(\vec r_i)S(\vec r_{i+1})}b_ib_{i+1} +
{\rm h.c.} \right)
\end{eqnarray}
In the long-wavelength limit, we replace the geometric mean
$\sqrt{S(\vec r_i)S(\vec r_{i+1})}$ by the arithmetic mean
$(S(\vec r_i)+S(\vec r_{i+1}))/2$, and then perform the Fourier
transform to the variables $b_i^\dagger$ and $b_i$. Notice that
the sites of classical spin pointing up and down are merged
together in their Fourier modes. Then
\begin{eqnarray}
 H_J &=& E_0 + 2S_0 J \sum_{\vec k} \left(
 2 b_{\vec k}^\dagger b_{\vec k} + \gamma_{\vec k} b_{\vec
 k}b_{-\vec k} + \gamma_{\vec k} b_{-\vec k}^\dagger b_{\vec
 k}^\dagger \right) \nonumber \\
 && + S_1 J \sum_{\vec k} \left( 2 b_{\vec k+2 \vec q_0}^\dagger b_{\vec k}
 + \gamma_{\vec k -2\vec q_0} b_{\vec k} b_{-\vec k + 2\vec q_0}
 \right.
 \nonumber \\
 &&\left. +\gamma_{\vec k +2\vec q_0} b_{\vec k} b_{-\vec k - 2\vec q_0} +
 {\rm h.c.}
 \right)
\end{eqnarray}
where $\gamma_{\vec k}=(\cos(k_x)+\cos(k_y))/2$, and the terms
involving higher harmonics are neglected. The Hamiltonian is
quadratic and in principle can be straightforwardly diagonalized.
Since we are only interested in the excitation spectrum around
$\vec k = \pm 2\vec q_0$, write
\begin{eqnarray}
H_J|_{\vec k \simeq \pm 2\vec q_0} =
 \sum_{\vec k \simeq 2\vec q_0} \eta_+^\dagger M_+ \eta_+
 + \sum_{\vec k \simeq -2\vec q_0} \eta_-^\dagger M_- \eta_-
\end{eqnarray}
where $\eta_\pm=(b_{\vec k} \ b_{\vec k \mp \vec q} \ b_{-\vec k
}^\dagger \  b_{-\vec k \pm \vec q}^\dagger)^T$ and
\begin{eqnarray}
M_\pm = J \left(\begin{array}{cccc}
 4S_0 & 2S_1 & 4S_0 \gamma_{\vec k} & 2S_1 \gamma_{\vec k \mp 2 \vec q} \\
 2S_1 & 4S_0 &  2S_1 \gamma_{\vec k \mp 2 \vec q} & 4S_0 \gamma_{\vec k \mp 2 \vec q} \\
 4S_0 \gamma_{\vec k} & 2S_1 \gamma_{\vec k \mp 2 \vec q} & 4S_0 & 2S_1 \\
 2S_1 \gamma_{\vec k \mp 2 \vec q} & 4S_0 \gamma_{\vec k \mp 2 \vec
 q} & 2S_1 & 4S_0
\end{array} \right) \nonumber \\
\end{eqnarray}
Note that it is the bosonic version of the spin-density wave
induced by charge stripes instead of the fermionic one due to
Fermi surface instability \cite{gruner}.

Diagonalize the Hamiltonian, we get the gapless excitation at
$\vec k = \pm 2\vec q_0$, with the anisotropic spin-wave
velocities, in which
\begin{eqnarray} \label{ekx}
\epsilon_{\vec k=(k_x,0)} = 4S_0 J \left[1 -
\frac{S_1^2}{4S_0^2}\sec^2(q_0)\right]^\frac{1}{2} |k_x \pm 2 q_0|
\end{eqnarray}
and
\begin{eqnarray} \label{eky}
\epsilon_{\vec k=(\pm 2 q_0,k_y)} = 4S_0 J |k_y|
\end{eqnarray}
One can estimate the ratio of the anisotropic spin-wave velocities
by the above Eqs.(\ref{ekx})-(\ref{eky}),
\begin{eqnarray} \label{vyvx}
\frac{v_x}{v_y} = \left[1 -
\frac{S_1^2}{4S_0^2}\sec^2(q_0)\right]^\frac{1}{2}
\end{eqnarray}
For LSCO, $q_0= 2\pi\delta$. $v_x/v_y=1$ at $\delta=0$. At $\delta
=1/8$, $v_x/v_y=0.9997$. Without the structural effect, one can
safely ignore the anisotropy of spin-wave velocities. However,
perturbed structural effect can be straightforwardly considered by
including a term
\begin{eqnarray}
- V \sum_i \cos(2 \vec q_0 \cdot \vec r_i)b_i^\dagger b_i
\end{eqnarray}
where $V > 0$ enhances the spin ordering of magnitude of period
$2R$. The term behaves as a perturbed term $\delta M$ to enhance
the first harmonic component $S_1$ of the spin magnitude.
\begin{eqnarray}
\delta M = \frac{-V}{2} \left(\begin{array}{cc}
 \begin{array}{cc}
 0 & 1 \\
 1 & 0
 \end{array} & 0 \\
 0 & \begin{array}{cc}
 0 & 1 \\
 1 & 0
 \end{array}
\end{array}\right)
\end{eqnarray}

For perturbed $V \ll t$ where the weak coupling limit is still
valid, the density profile $\rho(x)$ in Eq.(\ref{rho}) does not
vary. Eqs.(\ref{ekx}) and (\ref{eky}) are just modified by
replacing $S_1$ by $S_1 + V/(4J)$. At $\delta \ll 1$, we can
estimate
\begin{eqnarray} \label{vyvx2}
\frac{v_x}{v_y} \simeq \sqrt{1-\frac{V^2}{4J^2}}
\end{eqnarray}
The anisotropy in the neutron scattering measurement is around
0.75 \cite{aeppli}. Eq.(\ref{vyvx2}) gives an estimate of $V
\simeq 1.3J$.

If the structural effect is strong enough ($V/t \gtrsim 1$), for
example, in the sample of the ordered stripe phase of $\rm
La_2NiO_{4+\delta}$, we should go beyond the weak coupling limit
such that the density profile found in Eq.(\ref{rho}) should also
depend on $V$. The case of the ordered stripe phase will be
reported elsewhere.





The author would like to thank T.K. Lee for introducing him to the
stripe problem, and acknowledges numerous discussion with the
colleagues at the Institute of Theoretical Physics, Chinese
Academy of Sciences (ITP, CAS). The work was supported by the
National Science Council of Taiwan under Grant No.
NSC89-2816-M-001-0012-6, and the Visiting Scholar Program of ITP,
CAS under 20C905.



\newpage

\vspace{15pt}



\vspace{15pt}



\begin{thebibliography}{10}
\bibitem{otherHam}
Ref.[10] considered the hole stripes fluctuation inside the
three-leg ladder, and it is somehow equivalent to introducing the
structural effect implicitly.
\end{thebibliography}
\end{document}